# Cost Overruns and Demand Shortfalls in Urban Rail and Other Infrastructure

**Draft 6.0**

Bent Flyvbjerg

Aalborg University, Dept. of Development and Planning

Fibigerstraede 13, 9220 Aalborg, Denmark

Tel: +45 9635 8379, Email: flyvbjerg@plan.aau.dk

Bent Flyvbjerg is Professor of Planning with the Department of Development and Planning, Aalborg University, Denmark. He is founder and director of the university's research program on large infrastructure projects. His latest book is *Megaprojects and Risk* (with Nils Bruzelius and Werner Rothengatter). He has served as adviser on urban rail and other large infrastructure projects in the UK, Ireland, The Netherlands, Sweden, South Africa, and Denmark.

# Cost Overruns and Demand Shortfalls in Urban Rail and Other Infrastructure

**Draft 6.0: Please do not quote, copy, or circulate (except to referees)**

## Abstract

Risk, including economic risk, is increasingly a concern for public policy and management. The possibility of dealing effectively with risk is hampered, however, by lack of a sound empirical basis for risk assessment and management. The paper demonstrates the general point for cost and demand risks in urban rail projects. The paper presents empirical evidence that allow valid economic risk assessment and management of urban rail projects, including benchmarking of individual or groups of projects. Benchmarking of the Copenhagen Metro is presented as a case in point. The approach developed is proposed as a model for other types of policies and projects in order to improve economic and financial risk assessment and management in policy and planning.

**Keywords**: Urban rail, cost underestimation, cost overrun, cost escalation, demand overestimation, demand shortfalls, economic risk assessment and management, transportation infrastructure finance.

# The Age of Megaprojects

We live in the age of megaprojects. Never have so many expensive, large-scale projects been built over so short a historical period (Altshuler and Luberoff, 2003; Flyvbjerg, Bruzelius, and Rothengatter, 2003). Projects have grown larger over time, and increased size implies higher economic risks. A recent survey of top management in 25 of the largest construction firms in the world showed that executives see managing and pricing risk as one of their key challenges, and 63 percent of respondents said it was their biggest issue. Executives cited poor forecasting, poor risk identification, and cost escalation as the three top reasons for reduced profit margins (KPMG 2005). The basic problem is that reliable knowledge of risks is wanting or non-existent for most types of projects (Flyvbjerg, Holm, and Buhl, 2002, 2005; Morris and Hough, 1987; Szyliowicz and Goetz, 1995; World Bank, 1994). In the case of transportation infrastructure projects--rail, bridges, tunnels, roads, airports, seaports, terminals--studies of cost risks even point in opposite directions, with some authors (Pickrell, 1990) claiming that projects are highly risky, whereas others (Nijkamp and Ubbels, 1999) say this is not the case. A survey of the data and samples behind the studies explains why such diverse conclusions may exist side by side and why knowledge of risk is deficient: Previous studies are small-N research. Samples therefore cover too few projects to allow systematic, statistical analysis, and results of the studies are likely to depend on random properties of the selected samples (Flyvbjerg, Holm, and Buhl, 2002, 2005; Fouracre, Allport, and Thomson, 1990; Hall, 1980; Kain, 1990; Richmond, 1998; Walmsley and Pickett, 1992).

This conclusion holds, too, for the subject of the present paper, which is urban rail, a particularly costly type of public works project. Only few studies exist of the economic risks of urban rail projects and they are all small-N. The two best executed and most prominent studies in the field cover ten and 13 projects, respectively (Pickrell, 1990; Fouracre, Allport, and Thomson, 1990).[i] In comparison, the sample of urban rail projects established for the present study consists of 44 urban rail projects, which are compared with 214 other transportation infrastructure projects. The sample is the largest of its kind, and it allows, for the first time, statistically significant conclusions regarding the economic risks involved in building urban rail projects. The sample also allows statistically significant comparisons with other types of projects and between different geographical regions.

Absent or inadequate risk assessment and management are, in themselves, an important source of risk for projects. Because, until now, no reliable measure has been available for estimating risk in urban rail projects, effective risk assessment and management have been impossible. The study described below is aimed at changing this situation. It denotes a first step toward empirically grounded and valid risk assessment and management of urban rail projects by presenting and analyzing data that allow such risk assessment and management.

In addition to lack of a sound empirical basis, a main cause of absent risk assessment and management is lack of institutional checks and balances that would enforce accountability with rail project promoters towards risk. Such accountability would generate a demand for knowledge about real risks that is often absent today. The study described below documents a dire need for checks, balances, and accountability of this type. The work with developing procedures and institutional designs that, if implemented, would strengthen accountability towards risk has been begun elsewhere and will not be taken up here (Bruzelius, Flyvbjerg, and Rothengatter, 1998; Flyvbjerg, Bruzelius, and Rothengatter, 2003; Flyvbjerg and Cowi, 2004; Flyvbjerg, Holm, and Buhl, 2002, 2005). Finally, it should be stressed that absent risk assessment is not caused by lack of relevant methods. The methods exist and are fairly well developed, technically speaking. The problem is one of application. First, if methods of risk assessment are applied at all, applications are typically based on hypothetical, subjective data. This is due to the lack of empirical knowledge about risk mentioned above. Second, applications often have little or no bearing on real decision making, because of their lack of institutional grounding.

The following interesting questions are not addressed in the present paper, because they have been covered elsewhere: Why urban rail projects differ from other projects, the causes of cost underestimation and demand overestimation, possible differences between public and private projects, and how risk assessment and management may be designed in practice. The reader is referred to Flyvbjerg, Holm, and Buhl (2002, 2004, 2005), Flyvbjerg, Bruzelius, and Rothengatter (2003), and Flyvbjerg and Cowi (2004) for answers to these questions and for further references. Previous papers have covered costs and demand separately (Flyvbjerg, Holm, and Buhl, 2002, 2005). This paper brings the two together for the first time and focuses the analysis on urban rail and risk.

## Methods

"Urban rail" is here defined as rail in an urban area, including both heavy and light rail, which may be underground, at level, or elevated. "Risk" in urban rail projects is defined as downside uncertainty regarding costs and ridership. Urban rail projects are compared to other types of transportation infrastructure projects in order to test for differences. For a more detailed description of the methods involved than that given below, see Flyvbjerg (2005) and Flyvbjerg, Holm, and Buhl (2002, 2005).

Cost data for all projects consist of information about the difference between forecasted and actual construction costs for a sample of 258 transportation infrastructure projects. The project portfolio is worth approximately US$110 billion (2005 prices). In addition to urban rail, the portfolio includes bridges, tunnels, roads, high-speed rail, and conventional rail. The construction costs of projects range from US$1.5 million to 8.5 billion, with the smallest projects typically being stretches of roads in larger road schemes and the largest projects being rail links, tunnels, and bridges. The projects are located in 20 countries on five continents, including both developed and developing nations. 61 projects are North American, 181 European, and 16 are placed in other countries (developing nations and Japan). The projects were completed between 1927 and 1998. Older projects were included in the sample in order to test whether the accuracy of estimated costs improve over time.

Cost data for urban rail consist of information about the difference between forecasted and actual construction costs for a sample of 44 urban rail projects worth approximately US$37 billion (2005 prices). The 44 urban rail projects is a subset of the 258 projects mentioned above. Some of the rail projects are extensions of already existing systems. Of the 44 urban rail projects, 18 are located in North America, 13 in Europe, and 13 in developing nations. The projects were completed between 1966 and 1997.

For all projects the difference between forecasted and actual construction costs is calculated as actual costs minus forecasted costs in percent of forecasted costs. Forecasted costs are defined as estimated costs at the time of decision to build. This baseline is the international standard for calculating cost development. Actual costs are defined as real, accounted costs determined at the time of completing a project. All calculations are in constant prices. Thus a positive figure of, say, 25 indicates a cost escalation of 25 percent in constant prices. A negative figure similarly indicates a cost saving of that amount. Zero indicates that forecasted costs were correct and thus equal to actual costs.

Traffic data for all projects consist of information about the difference between forecasted and actual traffic for a sample of 210 transportation infrastructure projects. The projects again include, in addition to urban rail, bridges, tunnels, roads, high-speed rail, and conventional rail. The projects are located in 15 countries on five continents, covering both developed and developing nations. The projects were completed between 1969 and 1998.

Traffic data for urban rail consist of information about the difference between forecasted and actual traffic for 24 urban rail projects for which such data were available. Traffic is measured as number of passengers using the rail project in question, measured either as number of passengers per year or average daily ridership. The 24 urban rail projects is a subset of the 210 projects mentioned above. Of the 24 urban rail projects eight are in North America, six in Europe, and ten in developing nations.

For all projects the difference between forecasted and actual traffic is calculated as actual traffic minus forecasted traffic in percent of forecasted traffic. A completely accurate traffic forecast registers as zero. A positive figure of, say, 15 indicates actual traffic for a project was 15 percent higher than forecasted traffic, whereas a negative figure indicates actual traffic was that much lower than forecasted. Traffic is forecasted and counted for the opening year or the first full year of operations. The basis for calculating the difference between forecasted and actual traffic is the forecast at the time of decision to build the project. Again this baseline for calculations is the international standard.

The samples include all projects of the types mentioned for which reliable data were available. As far as the author knows, the samples are the largest of their kind and they allow for the first time statistically significant conclusions regarding the cost and revenue risks involved in the building of transportation infrastructure. If the samples are biased the bias is most likely conservative. Thus for costs, the sample is probably underestimating actual cost escalation in the project population. For traffic, the accuracy of forecasts estimated on the basis of the sample is probably higher than forecasts in the project population.

For all data, 25, 50, and 75 percent fractiles--also called quartiles--have been calculated. The lower quartile indicates 25 percent of data have a lower value than that indicated and 75 percent a higher, whereas for the upper quartile 75 percent of data have a lower value and 25 percent a higher. For instance, for rail in Table 1, 25 percent of projects have cost escalations of 24 percent or lower, 50 percent of projects have escalations of 43 percent or lower, and, finally, 75 percent of projects have escalations of 60 percent or lower. In addition, averages and standard deviations have been calculated. In comparing groups of data, one-sided variance analysis has been used with F-test, except for two groups where Welch's t-test was used. Theoretically, the statistical tests are based on normal distributions of data but the tests are quite robust against deviations from this. In one critical case the tests were supplemented with a non-parametric test.

## Cost Escalation in Transportation Infrastructure Projects

Table 1 shows the difference between forecasted and actual costs in 258 transportation infrastructure projects.

[Table 1 app. here]

Statistical analysis of the figures in the table show that means and standard deviations are significantly different for different project types. Projects, therefore, should not be pooled; each project type should be considered separately. The table shows:

- Rail has the largest cost escalation with an average of 44.7 percent, followed by bridges and tunnels with 33.8 percent, and finally roads with 20.4 percent.

- For rail, 75 percent of all projects have cost escalations of at least 24 percent. 25 percent of projects have cost escalations of at least 60 percent.

- The hypothesis that type of project has no effect on cost escalation is rejected at a very high level of statistical significance (p<0.001).

- The hypothesis that the error of underestimating costs is as common as the error of overestimating costs, or is numerically of the same size, is rejected with very high significance (p<0.001).

- For a randomly selected project the frequency of cost escalation is 86 percent whereas the frequency of correct forecasts or cost savings is 14 percent.

The size and frequency of cost escalation shown here, and the large standard deviations for forecasts of costs, documents a high level of uncertainty and risk as regards construction costs for transportation infrastructure projects in general, and particularly for rail, bridges, and tunnels.

## Cost Escalation for Urban Rail

Table 2 compares urban rail projects in three geographical areas: Europe, North America, and other countries. A test of differences between the three areas gives p=0.227. Thus there is no indication of difference; numerical variations can be considered random and projects in the three areas may be analyzed together.

[Table 2 app. here]

Table 3 compares urban rail to other rail (high-speed rail and ordinary rail) for which data were available regarding cost escalation. There is no significant difference between high-speed rail and ordinary rail (p=0.326). The two are therefore pooled and treated as one group under the heading "other rail." A test of difference between urban rail and other rail gives p=0.953. Thus there is no indication of difference but quite the opposite: The similarities are very high.

[Table 3 app. here]

Because there is no significant difference between urban and other rail the key question becomes how cost escalation for rail compares with cost escalation for other projects. And this question has already been addressed (see Table 1 and accompanying text).

For urban rail as well as other rail the conclusion is that these projects have cost escalations that are particularly large. For urban rail 75 percent of projects have cost escalations of at least 33 percent. 25 percent of urban rail projects have cost escalations of at least 60 percent.

For urban rail and other rail, large cost escalations combined with large standard deviations result in a particularly high level of uncertainty and risk regarding forecasts of costs, that is, budgets. The economic risk for such projects, here the risk that a given project turn out substantially more expensive than said at the time of making the decision to build the project, is significant. Assessment and management of such risk should therefore be central to all phases of the project development cycle in urban and other rail projects, from decision making to planning to construction.

## Demand Shortfalls in Transportation Infrastructure Projects

Traffic data consist of ridership for rail projects, including rail on bridges and in tunnels, and number of vehicles for road projects, including roads on bridges and in tunnels. Table 4 shows percentage differences between forecasted and actual traffic in the 210 transportation infrastructure projects in the sample.

[Table 4 app. here]

Statistical analysis of the figures in Table 4 shows the averages are significantly different across project types. Thus project types should be handled separately. The figures in the table show:

• For rail, actual traffic is on average 39.5 percent lower than forecasted traffic. For road, actual traffic is on average 9.5 percent higher than forecasted.

- For rail, 75 percent of projects have actual traffic that is at least 25 percent lower than forecasted traffic. 25 percent of projects have actual traffic that is at least 70 percent lower than forecasted.

- The upper and lower decil for rail (not shown in Table 4) show that only ten percent of projects achieve the traffic forecasted or more, whereas the lower ten percent of projects achieve 20 percent or less of forecasted traffic. For roads the figures are substantially more balanced.

- A statistical test of whether rail and road are different gives p<0.001. The difference is highly significant.

In sum, forecasted traffic is rarely achieved for rail while this is much more common for roads, even though a large standard deviation for roads indicates that for these projects, too, uncertainty and risk are high for traffic forecasts. For rail, large and frequent overestimations of ridership combined with a high standard deviation documents that uncertainty and risk are very high for traffic forecasts for this type of project.

## Demand Shortfalls for Urban Rail

Table 5 compares urban rail projects in three geographical areas: Europe, North America, and other countries.

[Table 5 app. here]

A test of difference between geographical areas gives p=0.004. Thus the difference is highly significant. The forecasts are more balanced for Europe than for the two other geographical areas even if the standard deviation is very high. However, two German urban rail projects have strongly diverging figures, namely 158 and 60 percent more passengers than forecast. If these two projects are treated as statistical outliers, as the distribution of data indicate should be the case, then the figures in Table 6 result.

A test of difference between geographical areas now gives p=0.054. The difference is no longer significant, even if the p-value is only just above the conventional critical value of 0.05. With a Kruskal-Wallis test, which is non-parametric and thus disregards normal distribution, p=0.074, that is, also non-significant. On this basis, no division between geographical areas is necessary in the statistical analysis. This conclusion is subject to the usual reservations for statistical analysis based on small numbers, especially for Europe. With only four observations and a p-value that is almost significant there is a marked need for information about more projects to ensure more valid conclusions as regards geographical subgroups of urban rail projects.

[Table 6 app. here]

For all urban rail projects, excluding the two German outliers, the following applies:

•       Actual ridership is on average 50.8 percent lower than forecast.

•       Only two projects out of 22 achieved the forecasted ridership.

•       75 percent of projects achieved a ridership that was at least 40 percent lower than forecast.

•       25 percent of projects achieved a ridership that was at least 68 percent lower than forecast.

In sum, for urban rail projects forecasted ridership is routinely far from achieved. Low actual ridership combined with a high standard deviation show that uncertainty and risk are very high for ridership forecasts for urban rail.

To the extent that ridership is the basis for revenues, which is almost always the case, then the high risk regarding ridership translates into an equally high economic risk. The figures show this risk should be taken very seriously in urban rail projects and should occupy a central place in preparing, deciding, and operating such projects.

## The Double Risk of Urban Rail

The analysis of construction costs show that urban rail projects on average turn out substantially more costly than forecast. At the same time the analysis of ridership show urban rail to achieve considerably fewer passengers than forecasted and thus lower revenues. Urban rail is therefore economically risky on two fronts, both as regards costs and as regards revenues. Urban rail is doubly risky and the possibilities for financing cost escalations incurred during construction through increased revenues from more passengers during operations will often be limited.

In order to analyze the double risk of urban rail in a more systematic fashion, all urban rail projects were identified for which data were available both for the difference between forecasted and actual costs and for the difference between forecasted and actual ridership. This is 14 projects, of which eight are located in North America and six in Europe.

Table 7 shows the data for the 14 projects. The double risk with both cost escalation and lower-than-forecasted ridership is exceedingly clear for these projects. However, the two German projects mentioned above, which should be considered statistical outliers, are included in the 14 projects in Table 7.

[Table 7 app. here]

Table 8 shows the data if the two German projects are excluded as statistical outliers, as they should be. The double risk is now even more pronounced with an average cost escalation of 40.3 percent combined with an actual ridership that is on average 47.8 percent lower than forecasted.

[Table 8 app. here]

With only 12 observations reservations must be made for small numbers. Yet, the numbers are so significant and are supported so distinctly by the larger number of observations in other parts of the analysis that the conclusion stands firm that urban rail projects are high-risk ventures because revenue risks amplify cost risks and create projects that are risky to the second degree.

## Benchmarking Urban Rail: the Case of the Copenhagen Metro

With a point of departure in the analysis above, it is possible to benchmark economic risk in specific urban rail projects in an empirically and statistically valid manner. Other rail projects, bridges, tunnels, roads, airports, seaports, and terminals can be benchmarked similarly. As an example of benchmarking in urban rail, this section focuses on the Copenhagen Metro in Denmark, which is currently under construction. The first two stages opened in 2002 and 2003, respectively, and a third stage is planned to open in 2007. The combined length of the first three stages is 21 kilometers with 22 stations; ten kilometers are in tunnel, three kilometers elevated, and eight kilometers at-grade. A 14-kilometer fourth stage is in the decision making phase.

In Denmark, Parliament typically approves large transport infrastructure investments by passing a law, which in addition to the approval contains specific terms of the investment in question. So too for the Copenhagen Metro. When in 1992 the Danish Parliament decided to authorize the first three stages of the metro, legislators specified main alignment of track, location of major stations, and indicated which parts would be built in tunnel, elevated, and at-grade, respectively. The Parliament also decided that the public companies, which would be responsible for each of the three stages, would be allowed to build "different types of light rail, including urban rail [tram], mini-metro, and magnetic rail" (Danish Parliament, 1992, §7). As a basis for the Parliament's decision, the Minister of Finance presented MPs with a cost estimate of DKK 3.11 billion covering "the transport infrastructure mentioned in the proposed law," of which DKK 2.91 billion was for the first three stages of the metro and DKK 0.2 billion for a road and bike routes (Danish Parliament, 1991a, 7; 1990 prices).[ii] The estimated number of passengers per year was 35 million for a tram and 43 million for a mini-metro; the latter would run more frequently than a tram and have shorter travel times (Danish Parliament, 1991b, question no. 19). The project would be financed by two sources of revenue: passenger incomes and incomes from the selling of public land, which would be serviced by the new rail system. Financing was backed by sovereign guarantees, i.e., if the project proved non-viable the taxpayer would pick up the bill.

In 1993, Ørestadsselskabet--the public company in charge of project development--was established. The company analyzed different types of urban rail and in 1994 estimated that a tram would cost DKK 3.9 billion and generate 47 million passengers per year, whereas a more advanced light rail system would cost DKK 4.9 billion and generate 77 million passengers per year, while, finally, an automated mini-metro would cost DKK 5.2 billion and generate 88 million passengers per year (National Audit Office of Denmark, 2000, 42, 1994 prices). On that basis, in October 1994, Ørestadsselskabet decided to build an automated mini-metro. The higher cost of the mini-metro would be justified by higher revenues from more passengers, or so Ørestadsselskabet reasoned. After tendering, construction started in 1996 and soon after the project incurred significant delay and further cost increases. In 2005, with the project 95 percent completed, costs had increased 157 percent in constant prices compared with the budget presented to Parliament in 1991-92; compared with the budget estimated by Ørestadsselskabet in 1994 the cost increase is 66% (Ørestadsselskabet, 2005, 29). Final costs will not be know until construction is completed in 2007 and contractor claims settled, which is expected to happen in 2008-09.

In 2002, when the first stage of the Copenhagen Metro opened to passengers, a main concern was whether ridership would live up to expectations. As a consequence of the cost escalations described above, high passenger revenues were crucial to project viability. In 2003, the first full year of operations, 20 million passengers rode the metro, which was 35% below the 31 million passengers forecast by Ørestadsselskabet (2004, 16) for that year. In 2004 ridership was 34 million, or 46% below the forecast, and in 2005 it was 36 million, or 41% below the forecast (Ørestadsselskabet, 2005, 10, 12; Ørestadsselskabet, 2006; National Audit Office of Denmark, 2004, 26). As a result of the lower-than-expected ridership, Ørestadsselskabet reduced its forecasts. For 2005, the forecast was 61 million passengers, which was lowered to 44 million and was not achieved. For 2008, the forecast was 73 million, lowered to 63 million. Finally, for 2010, 80 million passengers was lowered to 74 million (National Audit Office of Denmark, 2004, 26). The forecast on the basis of which Parliament approved the metro in 1992 cannot be directly compared with actual ridership, because the forecast assumes that all three stages of the metro would be operating by 2000 and none were operating this year, and in 2005 only two stages were in operation. But depending on the assumptions made one finds that ridership for the first full year of operations was approximately 40 percent lower than forecasted ridership. It may be argued, however, that since the higher passenger forecast made by Ørestadsselskabet in 1994 was used to justify the more expensive investment in an automated mini-metro, as mentioned above, this is the forecast that should be used as the basis for comparison. For 2004, the first year in which both stage 1 and stage 2 were in full operation, such a comparison can be made and it shows that actual ridership was 44 percent lower than that forecasted, which is equivalent to the forecast being overestimated by 79 percent (National Audit Office of Denmark, 2004, 26). In conclusion, the high ridership forecasts, which were used to justify investment in an expensive automated mini-metro, have not as yet materialized as real, paying passengers. In fact, actual ridership for the mini-metro is below that forecast by Ørestadsselskabet for two less expensive tram and light rail schemes, which were rejected by Ørestadsselskabet because these schemes would carry too few passengers (National Audit Office of Denmark, 2000, 42).

The cost escalations and demand shortfalls described above for the Copenhagen Metro have generated substantial interest and controversy among Danish lawmakers, citizens, media, and professionals. Furthermore, the ballooning budget and lower-than-forecasted ridership has resulted in two audits by the Auditor General of Denmark (2000, 2004). The Copenhagen Metro is the first metro in Denmark and people, including the auditors, have asked, for instance, whether cost escalations and demand shortfalls are common for this type of project? Whether escalations and shortfalls of the size encountered in Copenhagen are common? Whether the project is worse or better on these points than urban rail in other countries?

Such questions can now be answered in a statistically valid manner because a standard of comparison has been established with the data presented in the previous sections. For instance, the answer to the question of whether cost escalation is common for urban rail is unambiguously positive: 86 percent of the urban rail projects covered in the study above had cost escalations. However, a cost escalation of 157 percent, like that found for the first three stages of the Copenhagen Metro, is unusually high. Compared to the average cost escalation for all urban rail projects of 45 percent, which may be used as an "average practice" benchmark, the cost escalation in Copenhagen is more than three times higher (see Table 9). Even if we consider cost escalation only after the decision to build the automated mini-metro in 1994, and thus give the Copenhagen Metro a large advantage over the benchmark, the Copenhagen Metro still performs 21 percentage points poorer than the benchmark in terms of cost escalation. Among the 44 urban rail projects covered by the cost study above, only in developing nations do we find a project with cost escalation higher than that found in Copenhagen; all other projects have lower escalations.

[Table 9 app. here]

The answer to the question of whether demand shortfalls are common for urban rail projects must also be answered in the affirmative. 91 percent of the urban rail projects covered by the study above had demand shortfalls. The demand shortfall for the Copenhagen Metro is 6-11 percentage points lower than demand shortfall in the benchmark (see Table 9). Thus the Copenhagen Metro has performed better than the benchmark on this variable. However, when we combine the cost escalation of 157 percent with the demand shortfall of 40-45 percent, we begin to understand why the Copenhagen Metro is a project in deep financial trouble, and why it has attracted as much negative public attention as is the case.

It is telling of just how widespread and negative publicity has been that when a popular Danish money magazine recently asked its readers to choose the most wasteful person in Denmark among ten nominated for their misuse of citizens' money, the readers unambiguously chose the chairman of Ørestadsselskabet. The chairman was nominated for "having no control whatsoever over costs for the Metro, as regards both operating and construction costs, not to speak of ridership" (Penge og Privatøkonomi, no. 11, 2004; no. 1, 2005).

Such public concern over taxpayers' money is well founded, as may be demonstrated by considering development in the payback period for the Copenhagen Metro. When the metro was proposed to Parliament in 1991-92, the payback period was estimated at 14 years (Danish Parliament, 1991b, question 20). In 2005, the payback period had increased to 55 years (Ørestadsselskabet, 2005, 23). With a payback period that was now longer than the life span of several assets in the project, for instance rolling stock, reinvestments had to be considered, resulting in a payback period of 76 years, or 442 percent longer than originally estimated (Ørestadsselskabet, 2005, 24, 27).

Project viability has so far been based mainly on luck, in the sense that viability has been secured only by the lowest interest rate in recorded history and by an unprecedented real estate boom in Copenhagen. Sensitivity analyses carried out by Ørestadsselskabet (2005, 24) show that an increase in interest rate of one percentage point would make it impossible to pay back the project debt. The same holds for a 30 percent drop in sales prices of real estate, which partly fund the metro. In addition to these risks, there is the risk that passenger forecasts will continue not to live up to expectations and that Ørestadsselskabet may not be able to avoid paying the claims brought against it by its contractors. It is no wonder, then, that in 2005 the owners of Ørestadsselskabet--the Danish government and Copenhagen Municipality--decided to restructure the company in an attempt to arrive at a more viable set-up.

According to the Auditor General of Denmark (2000), one reason that financial risks in the Copenhagen Metro have so dramatically caught up with the project is lack of adequate risk assessment and management in the project. On this point, too, the Copenhagen Metro resembles urban rail projects in other parts of the world where sound risk assessment and management are also the exception rather than the norm. In retrospect, the single most important cause of troubles for the Copenhagen Metro is the decision made by Ørestadsselskabet in 1994 to build an expensive automated mini-metro, based on the expectation that the increased expense--compared with less expensive tram and light rail systems--would be more than offset by even larger increases in passenger revenues. The cost and revenue risks involved in this decision were severely underestimated. As a consequence, the already expensive mini-metro became even more expensive and the passengers that would justify the investment did not appear. Hopefully for Danish taxpayers, the planners and managers of the planned DKK 15 billion, all-tunnel fourth stage of the Copenhagen Metro will be better at managing risks--and at forecasting costs and ridership--than the planners and managers for the first three stages. And hopefully others, too, may learn from the very expensive lessons regarding risk assessment and management taught by the first three stages of the Copenhagen Metro.

## Economic Risk in Other Policy Areas

In addition to data on economic risk for transportation infrastructure projects, the author has reviewed such data for several hundred other projects in other policy areas, including power plants, dams, water projects, oil and gas extraction projects, information technology systems, aerospace projects, and weapons systems (Arditi, Akan, and Gurdamar, 1985; Canaday, 1980; Blake, Cox, and Fraize, 1976; Department of Energy Study Group, 1975; Dlakwa and Culpin, 1990; Fraser, 1990; Hall, 1980; Healey, 1964; Henderson, 1977; Hufschmidt and Gerin, 1970; Merrow, 1988; Morris and Hough, 1987; World Bank, 1994; World Bank, undated). The data indicate that other types of projects and other policy areas incur economic risks in the same order of magnitude as the risks involved in transportation infrastructure projects.

Among the more spectacular examples are the Sydney Opera House with actual costs approximately 15 times higher than those projected and the Concorde supersonic airplane with 12 times higher costs (Hall, undated, p. 3). The data also indicate that economic risks have neither increased nor decreased historically and that underestimation is common in both first and third-world countries (Flyvbjerg, Holm, and Buhl, 2002, 2005). When the Suez canal was completed in 1869 actual construction costs were 20 times higher than the earliest estimated costs and three times higher than the cost estimate for the year before construction began. The Panama Canal, completed in 1914, had cost escalations in the range from 70 to 200 percent (Summers, 1967, p. 148). In sum, the phenomenon of substantial economic risk appear to be characteristic not only of urban rail and other transportation projects but of projects in other policy areas as well.

## Summary and Discussion

Risk, including economic risk, is increasingly a concern for public policy and management. The possibility of dealing effectively with risk is hampered, however, by lack of a sound empirical basis for risk assessment and management. The paper demonstrates this point for cost and revenue risks in urban rail projects. The paper presents, for the first time, empirical evidence that allow valid economic risk assessment and management of such projects, including benchmarking of individual or groups of projects.

The paper shows that urban rail projects are particularly risky ventures, although other transportation projects, like tunnels and bridges, are also highly risky, as are projects in other policy areas than transportation:

- Average cost escalation for urban rail is 45 percent in constant prices.

- For 25 percent of urban rail projects cost escalations are at least 60 percent.

- Actual ridership is on average 51 percent lower than forecasted.

- For 25 percent of urban rail projects actual ridership is at least 68 percent lower than forecasted.

When cost risk and revenue risk are combined, a risk profile emerges for urban rail, which proves such projects to be economically risky to the second degree.

This conclusion is not intended to serve as an argument against building urban rail; many cities, undoubtedly, need urban rail to solve their transportation problems. Neither is it an attack on public--vs. private--spending on infrastructure, because the data do not allow an answer to the interesting question of whether private projects perform better or worse than public ones in terms of economic and financial risk. Finally, the data and conclusions do not warrant an attack on spending on transportation vs. spending on other policy areas, because projects in other policy areas appear as liable to cost escalation and lower-than-forecasted revenues as are transportation projects.

What the data and conclusions do establish--with urban rail as an in-depth case study--is that significant cost underestimation and revenue overestimation are widespread practices in project development and implementation. Such practices are sources of substantial risk and they form real barriers to the effective allocation of scarce resources for building important infrastructure. The data document a need for implementing institutional checks and balances that would change the current practices of cost underestimation and revenue overestimation to ones where empirically based and statistically valid risk assessment and management are effectively used in all phases of the policy and project development process, from planning to decision making to construction to operations, in order to properly identify, reduce, and manage risk.

With actual costs and patronage in urban rail being different from that forecasted to the degree and with the statistical significance documented above, an inevitable conclusion is that the results of conventional cost-benefit analysis, which is typically at the core of documentation and decision making for this type of project, is of little or negative value here. A cost-benefit analysis based on the forecasts of costs and revenues described above would, with a high degree of certainty, be strongly misleading. Garbage in, garbage out, as the saying goes. This does not show the uselessness of cost-benefit analysis as such, needless to say. But if informed decisions are the goal, then empirical risk analysis must supplement conventional cost-benefit analysis as the main means for documenting and deciding on urban rail projects.

Given the data presented above, a key policy recommendation for legislators and citizens who care about what Williams (1998) calls "honest numbers" is they should not trust the budgets, patronage forecasts, and cost-benefit analyses produced by project promoters and planners of urban rail. Independent studies should be carried out, and, again, such studies should be strong on empirically based risk assessment and management.

Until now it has been difficult or impossible to carry out meaningful economic risk assessment and management for urban rail projects, because empirically grounded and statistically valid figures of risk did not exist for this type of project. With the study documented above such figures now exist and empirical risk assessment and management can begin. In addition to sound data, institutional checks and balances that would enforce accountability in actors towards risk are also necessary to make risk assessment and management work. The labor with developing such checks and balances has been begun elsewhere (Bruzelius, Flyvbjerg, and Rothengatter, 1998; Flyvbjerg, Bruzelius, and Rothengatter, 2003; Flyvbjerg and Cowi, 2004; Flyvbjerg, Holm, and Buhl, 2002, 2005).

Two areas stand out as particularly pertinent for further research. One is further data collection, the other is development and test of explanatory models that would clarify the peculiar bias found in cost and traffic forecasts for urban rail. As regards further data collection, even though the sample used in the study is the largest of its kind, it is too small to allow more than the crudest subdivisions, and thus the simplest of comparisons and explanations. More data on more urban rail schemes are needed in order to see whether different types of schemes perform differently, for instance light rail compared with heavy rail, driverless metros compared with ordinary metros, and rail above ground compared to underground rail. More geographical data are needed as well, especially for patronage in Europe, where the number of observations are particularly low, making comparisons between different parts of the world difficult. With such data risk assessment and management could be further improved.

As for explanatory models, when, as a researcher, you uncover a body of information for decision making that is as systematically and significantly biased as that documented above for urban rail--with strongly underestimated costs and strongly overestimated ridership--you obviously begin to ask what causes such striking bias and which purposes does it serve? These questions are particularly pertinent for projects as costly and consequential as urban rail projects, each of which often measures in the billion-dollar range. Playing the devil's advocate one might argue that urban rail investments are not risky at all. Calling them risky implies that there is some high uncertainty with respect to outcome. The evidence presented above suggests just the opposite, however. Outcomes are very certain. Ridership will almost certainly be lower than forecasted. Costs will almost certainly be higher than forecasted. There is a difference between making a risky investment and an investment that will almost surely underperform. The question is why the latter type of investment is repeatedly made in the field of urban rail, as if no learning takes place.

Work that attempts to answer such questions has recently been begun elsewhere by testing different types of explanatory models of bias and risk: technical, economic, psychological, and political. As part of this exercise tests are carried out of how data fit the possibility that urban rail forecasts are strategically and deliberately misleading, that is, that forecasted costs and revenues are produced, not as best estimates of what the future will bring if projects are built, but instead as misrepresentation constructed to get projects built (Flyvbjerg, Holm, and Buhl, 2005; Flyvbjerg, in progress).

## Acknowledgments

The author wishes to thank Mette K. Skamris Holm for help with data collection for this paper and Søren L. Buhl for carrying out statistical tests. The author also wishes to thank Bert van Wee, Hugo Priemus, and two anonymous referees for valuable comments on an earlier draft of the paper. The research for the paper was supported by Aalborg University and the Danish Transport Council.

*Table 1: Average cost escalation in 258 transportation infrastructure projects. Constant prices.*

| Project type | No. of projects (N) | Quartiles (25/50/75%) | Average cost escalation (%) | Standard deviation |
|---|---|---|---|---|
| Rail | 58 | 24/43/60 | 44,7 | 38,4 |
| Bridges and tunnels | 33 | -1/22/35 | 33,8 | 62,4 |
| Roads | 167 | 5/15/32 | 20,4 | 29,9 |
| All projects | 258 | 5/20/35 | 27,6 | 38,7 |

*Table 2: Average cost escalation in 44 urban rail projects in three geographical areas. Constant prices.*

| Project type | No. of projects (N) | Quartiles (25/50/75%) | Average cost escalation (%) | Standard deviation |
|---|---|---|---|---|
| Europe | 13 | 39/45/57 | 43,3 | 21,3 |
| North America | 18 | 33/42/54 | 35,8 | 30,4 |
| Other | 13 | 35/59/75 | 59,2 | 53,6 |
| All | 44 | 33/44/59 | 44,9 | 37,3 |

*Table 3: Average cost escalation in 58 rail projects. Constant prices.*

| Project type | No. of projects (N) | Quartiles (25/50/75%) | Average cost escalation (%) | Standard deviation |
|---|---|---|---|---|
| Urban rail | 44 | 33/44/60 | 44,9 | 37,3 |
| Other rail | 14 | 10/34/75 | 44,1 | 43,3 |
| All | 58 | 24/43/60 | 44,7 | 38,4 |

*Table 4: Difference between forecasted and actual traffic in 210 transportation infrastructure projects.*

| Project type | No. of projects (N) | Quartiles (25/50/75%) | Average difference (%) | Standard deviation |
|---|---|---|---|---|
| Rail (ridership) | 27 | -70/-54/-25 | -39,5 | 52,4 |
| Roads (vehicles) | 183 | -18/0/28 | 9,5 | 44,3 |
| All | 210 | -24/-4/24 | 3,2 | 48,2 |

*Table 5: Difference between forecasted and actual ridership in 24 urban rail projects in three geographical areas.*

| Project type | No. of projects (N) | Quartiles (25/50/75%) | Average difference (%) | Standard deviation |
|---|---|---|---|---|
| Europe | 6 | -29/-4/45 | 20,7 | 77,3 |
| North America | 8 | -69/-63/-53 | -60,0 | 17,0 |
| Other | 10 | -70/-57/-50 | -54,3 | 27,5 |
| All | 24 | -67/-53/-27 | -37,5 | 53,5 |

*Table 6: Difference between forecasted and actual ridership in 22 urban rail projects in three geographical areas. Two statistical outliers excluded.*

| Project type | No. of projects (N) | Quartiles (25/50/75%) | Average difference (%) | Standard deviation |
|---|---|---|---|---|
| Europe | 4 | -40/-22/-6 | -23,5 | 23,5 |
| North America | 8 | -69/-63/-53 | -60,0 | 17,0 |
| Other | 10 | -70/-57/-50 | -54,3 | 27,5 |
| All | 22 | -68/-54/-40 | -50,8 | 26,1 |

*Table 7: Cost escalation and ridership for 14 urban rail projects.*

|  | Quartiles (25/50/75%) | Avg. difference btwn. actual and forecasted development (%) | Standard deviation |
|---|---|---|---|
| Costs | 35/45/57 | 42,9 | 25,4 |
| Ridership | -65/-50/-13 | -25,4 | 64,6 |

*Table 8: Cost escalation and ridership for 12 urban rail projects. Two statistical outliers excluded.*

|  | Quartiles (25/50/75%) | Avg. difference btwn. actual and forecasted development (%) | Standard deviation |
|---|---|---|---|
| Costs | 28/45/56 | 40,3 | 25,3 |
| Ridership | -67/-52/-34 | -47,8 | 25,6 |

*Table 9: Benchmarking the Copenhagen Metro. The benchmark is defined as average practice among urban rail projects, based on samples of 44 and 22 projects for cost escalation and demand shortfalls respectively.*

| | Copenhagen Metro | Benchmark (average practice) | Diff. btwn. Copenhagen Metro and benchmark |
|---|---|---|---|
| Cost escalation % (constant prices) | 157 | 45 | 112 |
| Demand shortfall % | 40-45% | 51 | -11 to -6 |

# Notes

---

[i] A third study exists, which covers 17 urban rail projects (Merewitz, 1973a, b). This is, to our knowledge, the only earlier study of urban rail and transportation infrastructure with an attempt at statistical analysis. This study aimed at comparing cost overrun in urban rapid transit projects, and especially overrun in the San Francisco Bay Area Rapid Transit (BART) system, with overrun in four other types of public works projects. We have replicated this study and found that in addition to issues of small-N sampling, the handling of data raises a number of problems. First, cost data were not corrected for inflation, that is, current prices were used instead of constant prices. This is known to be a major source of error due to varying inflation rates between projects and varying duration of construction periods. Second, in statistical tests the mean cost overrun of subgroups of projects, for instance urban rail, was compared with the grand mean of overrun for all projects, thus making the error of comparing projects with themselves. Subgroups should have been tested directly against other subgroups in deciding whether they differ at all and, if so, which ones differ. Third, Merewitz (1973a) and (1973b) are inconsistent. Merewitz (1973a) calculates the grand mean of cost overrun as the average of means for subgroups; that is, the grand mean is unweighted where common practice is to use the weighted mean, as appears to be the approach taken in Merewitz (1973b). Fourth, due to insufficient information the p-values calculated are difficult to verify; most likely they are flawed, however, and the use of one-sided p-values is misleading. Finally, a debatable assumption about symmetry was used, which has more impact for the non-parametric test used than non-normality has for parametric methods. Despite these shortcomings, the approach taken in this study was innovative for its time and in principle pointed in the right direction regarding how to analyze risks from cost overrun in public works projects.

[ii] These and later prices do not include VAT.